%% file: conference_101719.tex
\documentclass[conference]{IEEEtran}
\IEEEoverridecommandlockouts
\usepackage{cite}
\usepackage{amsmath,amssymb,amsfonts}
\usepackage{algorithmic}
\usepackage{graphicx}
\usepackage{textcomp}
\usepackage{xcolor}
\def\BibTeX{{\rm B\kern-.05em{\sc i\kern-.025em b}\kern-.08em
    T\kern-.1667em\lower.7ex\hbox{E}\kern-.125emX}}

\usepackage{url}
\usepackage{hyperref}
\hypersetup{
    colorlinks=true,
    linkcolor=black,
    filecolor=magenta,      
    urlcolor=black,
    citecolor=black,
    }
    
\usepackage{color,soul}

\usepackage{listings}
\definecolor{codegray}{rgb}{0.5,0.5,0.5}
\definecolor{codegreen}{rgb}{0,0.5,0}
\definecolor{backcolour}{rgb}{0.95,0.95,0.95}

\lstdefinestyle{mystyle}{backgroundcolor=\color{backcolour},  commentstyle=\color{codegreen}, keywordstyle=\color{blue}, numberstyle=\tiny\color{codegray}, stringstyle=\color{green}, basicstyle=\ttfamily\tiny, breakatwhitespace=false, breaklines=true, captionpos=b, keepspaces=true, numbers=left, numbersep=5pt, showspaces=false, showstringspaces=false, showtabs=false, tabsize=2}
\definecolor{mGreen}{rgb}{0,0.6,0}
\definecolor{mGray}{rgb}{0.5,0.5,0.5}
\definecolor{mPurple}{rgb}{0.58,0,0.82}
\definecolor{backgroundColour}{rgb}{0.95,0.95,0.92}

\lstdefinestyle{CStyle}{
    backgroundcolor=\color{backgroundColour},   
    commentstyle=\color{mGreen},
    keywordstyle=\color{magenta},
    numberstyle=\tiny\color{mGray},
    stringstyle=\color{mPurple},
    basicstyle=\footnotesize,
    breakatwhitespace=false,         
    breaklines=true,                 
    captionpos=b,                    
    keepspaces=true,                 
    numbers=left,                    
    numbersep=5pt,                  
    showspaces=false,                
    showstringspaces=false,
    showtabs=false,                  
    tabsize=2,
    language=C
}

\definecolor{light-gray}{gray}{0.9}
\lstdefinestyle{DOS}
{
    backgroundcolor=\color{light-gray},
    basicstyle=\scriptsize\color{black}\ttfamily,
    columns=fullflexible,
    numbers=none
}

\begin{document}

\title{CARM Tool: Cache-Aware Roofline Model Automatic Benchmarking and Application Analysis
}

\author{
\IEEEauthorblockN{José Morgado, Leonel Sousa and Aleksandar Ilic}
\IEEEauthorblockA{\textit{INESC-ID, Instituto Superior Técnico, Universidade de Lisboa} \\
Lisbon, Portugal \\
\{jose.a.morgado,leonel.sousa,aleksandar.ilic\}@inesc-id.pt}

}

\maketitle

\begin{abstract}
In recent years, HPC systems and CPU architectures as their central components have become increasingly complex, making application development and optimization quite challenging. To this respect, intuitive performance models like the Cache-aware Roofline Model (CARM) offer effective guidance by providing insights into bottlenecks that limit the application's ability to reach the system's maximum performance. To fully exploit the benefits of CARM optimization guidance for application development, automatic tools for cross-architecture model construction and in-depth application characterization are absolutely essential. Given a plethora of existing CPU architectures, the current landscape of CARM-enabled tools covers either vendor-specific (Intel Advisor), not sufficiently developed (ARM) or simply non-existing (AMD, RISC-V) tools. This is a particular gap that this work intends to close by bringing automatic CARM support to all major CPU architectures and ISAs, i.e., x86 (Intel, AMD), ARM, and RISC-V, by developing assembly microbenchmarks specifically tailored to cover a full performance spectrum of modern CPUs (from scalar to all supported vector ISA extensions) for both computational units and all memory hierarchy levels. 
Additionally, this work integrates application analysis within the CARM framework using performance counters and dynamic binary instrumentation.  
Experimental 
results show that the CARM roofs constructed with the proposed automated
framework provide less than a 1\% deviation across 
various tested architectural maximums.

\end{abstract}


\section{Introduction}
\input{introduction}
\section{Background and Related Work}
\input{state_of_the_art}
\section{CARM Tool: High-Level Overview}\label{CARM_Tool_high_level}
\input{CARM_Tool_high_level}
\section{CARM Tool: Low-Level Overview}
\input{CARM_Tool_low_level}
\section{Experimental Results}
\input{experimental_results}

\section{Conclusions}
\input{conclusion}

\section*{Acknowledgements}

This work was supported by European Union HE Research and Innovation programme under grant agreement No 101092877 (SYCLOPS), and FCT (Fundação para a Ciência e a Tecnologia, Portugal) through the UIDB/50021/2020 project. The research presented in this paper has benefited from the Experimental Infrastructure for Exploration of Exascale Computing (eX3), which is financially supported by the Research Council of Norway under contract 270053. We also thank 
Diogo Marques for his preliminary work in the development of the Intel microbenchmarks and the tool itself.

\newpage

\hypersetup{
    colorlinks=true,
    linkcolor=black,
    filecolor=magenta,      
    urlcolor=black,
    citecolor=black,
    }

\bibliographystyle{IEEEtranS}
\bibliography{reference}


\hypersetup{
    colorlinks=true,
    linkcolor=black,
    filecolor=magenta,      
    urlcolor=cyan,
    citecolor=black,
    }
\urlstyle{same}

\newpage

\appendix
\section{Artifact Appendix}
\input{artifact_appendix}
\end{document}

%% file: introduction.tex

As High Performance Computing (HPC) systems continue to grow in complexity and scale, their computational performance has also grown exponentially, allowing for the execution of increasingly demanding workloads \cite{articleHPC}. 
This complexity often comes in the form of computing platform heterogeneity, in the structure of multi-core Central Processing Unit (CPU)s, their execution pipelines, as well as in deeper memory subsystems, including diverse emergent technologies \cite{10.1145/3528535.3565238}.
As a consequence of this complexity, the understanding, profiling, and benchmarking of these systems have become far from a trivial task, which has led to increased difficulty in the development of applications that take full advantage of the system hardware resources \cite{9773216}. 
Another factor is the diversity of system architectures across different vendors (and even across CPU generations of the same vendor), necessitating distinct approaches to application optimization and resulting in the divergent performance of the same application on different architectures. 
Furthermore, the demands and hardware requirements of certain applications can vary during their execution, making it challenging to identify their performance bottlenecks \cite{marques2020application}.
For instance, load and store instructions may have different latencies on different architectures, and variations in Instruction Set Architecture (ISA) support on architectures can impact performance.

In these scenarios, architecture, and application performance analysis are often necessary to assess these limitations, typically via the use of performance models. To this respect, the roofline model is of particular interest~\cite{williams2009roofline} -- a performance model known for its easy-to-understand guidelines and useful insights into what bottlenecks are constraining application performance on a given system.
Starting from the Original Roofline Model (ORM)~\cite{williams2009roofline}, there have been several notable model improvements that include the Empirical Roofline Model (ERC)~\cite{8639946}, the Integrated Roofline Model (IRM)~\cite{10.1007/978-3-319-92040-5_12}, the Cache-Aware Roofline Model (CARM) ~\cite{ilic2013cache}, and the Instruction Roofline Model~\cite{9059264} among others.
These types of models have varying levels of difficulty in their implementation to be generated for a given system. For example, a fully-compliant ORM implementation requires measuring the memory traffic between cache levels
usually done via cache simulation,
which is very time-consuming and architecture-dependent.
On the other hand, the CARM requisites are less complex to be made portable across various architectures, since the CARM relies on microbenchmarking data for various Instruction Set Architectures (ISAs). 
In fact, the CARM is a widely used roofline model 
due to its ability to provide a more detailed look at an architecture by considering the diverse characteristics of the memory subsystem, which typically varies across different levels in terms of size, bandwidth, and latency. 
This approach results in some advantages over the ORM such as the ability to maintain a constant Arithmetic Intensity (AI) for different problem sizes (prevents inaccurate classification of applications), and more accurate characterization of the maximum performance and bandwidth limits of the system memory levels.

Currently, the CARM is supported for Intel architectures via Intel Advisor \cite{IntelAdvCARM}, while the ORM has rudimentary support in AMD via AMD uProf \cite{AMDManual} which are closed-source tools, while open-source tools like the Empirical Roofline Tool (ERT) \cite{ERT_bitbucket} are also available but their support is also mostly focused on x86-64 architectures, and does not provide accurate benchmarking mechanisms for architecture maximum performance exploration.
This work aims to close this gap in availability for different architectures such as AARCH64 and RISCV64 by developing an easy-to-use open-source portable tool, that is able to generate the CARM for various architectures based on 
automatically generated tailored 
assembly microbenchmarks and also provide application analysis in the context of CARM, with a Graphical User Interface (GUI) to further facilitate the visualization of results\footnote{The proposed open-source tool is available at  
\url{https://github.com/champ-hub/carm-roofline}}.



The main contributions of this work are as follows:

\begin{itemize}
    \item Establishing a cross-architecture microbenchmark methodology;

    \item The development of portable microbenchmarks to generate the CARM model for various mainstream CPU architectures, such as the most recent Intel, AMD, ARM, and RISC-V processors;

    \item The integration of performance counter and DBI analysis of applications to then examine in the scope of the CARM model;

    \item The implementation of these various features into a single tool, providing a one-stop shop for CARM-related analysis, with a GUI to facilitate usage

    \item Analysis of different architectures and applications using said tool, with promising results accurately reaching architectural limits.
\end{itemize}




%% file: state_of_the_art.tex



The landscape of roofline modeling features two predominant models: the CARM~\cite{ilic2013cache}, and the ORM~\cite{williams2009roofline}. While all models correlate the performance with Arithmetic Intensity (AI), they differ in their approach to memory traffic, which affects their performance characterization capabilities and 
leads to distinct insights into memory operation impact. 
The CARM observes the memory traffic from the core perspective, i.e., it considers all load and store operations and the complete memory hierarchy in a single plot, thus defining the true application AI. This  property allows extending its insightfullness by incorporating a set of features unique to CARM, such as 
application-specific load/store ratios, memory ports utilization, various instruction sets and data precision, thereby establishing more accurate performance upper-bounds that align with actual application requirements~\cite{marques2020application}. 
In contrast, ORM observes the traffic between two subsequent levels of memory hierarchy, thus it consists of several models (one for each memory level). As such,  ORM extensions incorporate micro-architectural throughput analyses through simulation, which may hinder its adaptability in diverse optimization frameworks. 



As evidenced in its full integration in 
Intel Advisor~\cite{IntelAdvCARM}, the CARM single-plot nature maintains simplicity and enhances usability through direct source-code and assembly analysis.
Its ability to correleate the capabilities of the entire memory hierarchy with 
peak compute throughput offers enhanced insights into the specific bottlenecks that impede application performance. 
As a consequence, the CARM brings to practice several advantages over the existing roofline approaches, which 
include the ability to identify the memory level requiring optimization priority for improved performance gains, the ability to maintain a constant AI for different problem sizes, and more accurate characterization of the maximum bandwidth of memory hierarchy levels. 

\begin{figure}[!t]
\centerline{\includegraphics[scale=0.6]{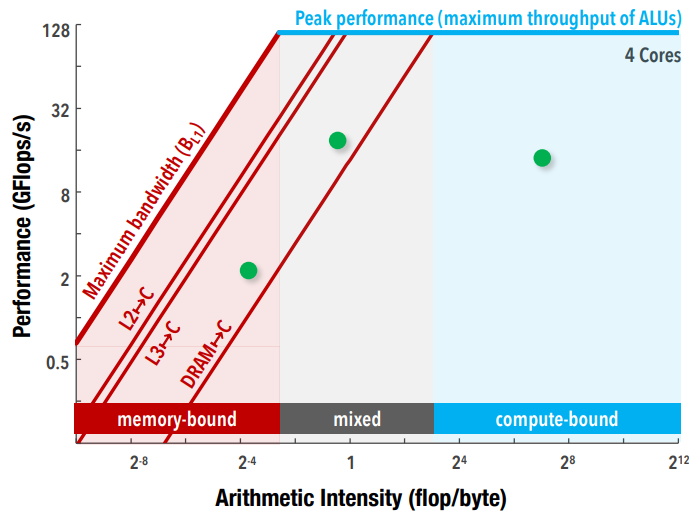}}
\caption{Cache-Aware Roofline Model with example kernels}
\label{CARMModel}
\end{figure}

To provide more detailed feedback for application bottlenecks and optimization steps, 
CARM considers all memory operations issued by the core, which may provoke accesses to different cache levels, thus served at different bandwidths. As a result, the CARM of a microarchitecture contains several sloped roofs, each discriminating the performance limitations imposed by the sustainable bandwidth of a specific memory level, as can be observed in Figure \ref{CARMModel}. The horizontal roof corresponds to the peak compute FP performance. From this plot, we can infer that there are three regions: memory-bound, mixed, and compute-bound regions, which are used to pinpoint the main culprits of application bottlenecks. An application in the memory-bound region should be first optimized to improve memory accesses, 
and  depending on which of the sloped roofs the application is plotted between, it becomes possible to identify the memory level that requires prioritized optimization. An application that falls under the compute-bound region should first prioritize improvements in the utilization of the CPU compute units, while an application in the mixed region can benefit from both types of improvements.

To build these roofs, different microbenchmarks must be used to evaluate and measure the bandwidth limits of all the memory levels and peak FP performance of the CPU. These assembly-level microbenchmarks are also validated using performance counters, and they are carefully designed to exploit the different theoretical maximums of the specific microarchitecture that they are being run on. To build the sloped bandwidth roofs, a microbenchmark that can vary the number of memory operations to hit different memory levels by accessing contiguous and increasing memory addresses is used. 
For the flat FP performance roof, a microbenchmark that maximizes occupation of the CPUs arithmetic units pipeline is developed. From these microbenchmarks, one can derive the peak bandwidth of different memory levels -- B\textsubscript{Lx\textrightarrow C}, where $x{\in}\{L1,L2,...,DRAM\}$, and peak FP performance (F\textsubscript{p}). By folowing the observation that the memory transfers and computations overlap in time, the CARM performance roofs are expressed as follows:
\begin{equation} \label{CARMequation}
\centering
F_{a} = min(F_{p}, B_{Lx->C}\times AI),
\end{equation}
where AI is defined as a ratio between the number of FP operations and the amount of bytes transferred. 

The need for the microbenchmarks for CARM generation makes it a non-trivial task to provide support for different CPU microarchitectures. 
This represents one of the main objectives of this work, i.e., to provide support for CARM via the development of specific microbenchmarks compatible with CPU microarchitectures from different vendors and ISAs, such as Intel, AMD, ARM, and RISC-V. 
To showcase the cross-platform architectural diversity and performance disparity, we provide the microarchitecture analysis in Table~\ref{TableCPU}  that offers a comparative overview of theoretical performance metrics across various CPU microarchitectures from different vendors, focusing on CARM-related parameters. 
In particular, we consider four distinct CPU microarchitectures: Intel Skylake X (SKL-X), AMD Zen 3, ARM Vulcan, and RISC-V XuanTie C920 (Sophon SG2042). For each, we analyse the theoretical peak L1 cache bandwidth , as well  as the peak FP operations per cycle for various SIMD capabilities and ISA extensions, such as SSE, AVX2, AVX-512, NEON, and RVV.

\begin{table}[t!]
\begin{center}
\caption{Theoretical CARM metrics for various CPU~\label{TableCPU}}
\begin{tabular}{|c|c|c|c|c|} 
 \hline
  & \textbf{SKL-X} & \textbf{Zen 3} & \textbf{Vulcan} & \textbf{C920}\\
 \hline
 L1 bandwidth (B/cycle) & 192 & 96 & 32 & 32\\ 
 \hline
 Scalar DP (FP/cycle) & 4 & 4 & 4 & 4\\
 \hline
 SSE/NEON/RVV DP (FP/cycle) & 8 & 8 & 8 & 8\\
 \hline
 AVX DP (FP/cycle) & 16 & 16 & X & X\\
 \hline
 AVX-512 DP (FP/cycle) & 32 & X & X & X\\
 \hline
\end{tabular}
\end{center}
\end{table}

This analysis is based on the theoretical hardware specifications of the architectures, e.g., the L1 bandwidth is calculated by analyzing the number of load/store units that each CPU contains to determine the maximum amount of load/store Instructions per Cycle (IPC) that these units are capable of delivering, and then multiplying this value by the width of the operand size for different SIMD extensions supported in each architecture. For example, Intel Skylake-X and AMD Zen 3 CPUs both contain three load/store units (delivering three load/store IPC), but, since the Skylake-X implements the AVX512 ISA extension with 64-byte operands, it should be able to achieve a 192 byte per cycle of L1 bandwidth, while the Zen3 can achieve 96 with AVX2.
Both ARM Vulcan and RISC-V C920 contain two load/store units, leading to an expected memory IPC of two, since both of these CPUs implement 16-byte operands in their widest ISA extension (Neon and RVV with 16-byte width) we can expect a bandwidth of 32 bytes per cycle.
Regarding the theoretical peak FP operations per cycle, this value can be obtained by examining the amount of fused multiply-add (FMA) capable FP units in each core, which in this case is two on all of the different architectures, resulting in an expected two FMA IPC which then leads to four FP operations per cycle. 
Then with each ISA extension of greater width, we can see this peak performance double, as each consequent ISA is capable of twice as many FP operations per instruction. 

However, it is often observed that the specifications provided by manufacturers referred to as "peak" bandwidths in Intel's documentation \cite{IntelManual} versus "sustained" bandwidths for example, do not always align with achievable performance in real-world applications. This discrepancy highlights the necessity for the development of microbenchmarks that can accurately determine real-world values, thereby facilitating the construction of a precise CARM. 

Existing tools equipped with automatic CARM generation are mostly limited to x86-64 architectures, via closed-source tools such as Intel Advisor~\cite{SDE}, and AMD uProf \cite{AMDManual}, of which the latter only implements a roofline construction based on theoretical specifications. Open-source tools such as ERT~\cite{ERT_bitbucket} are also available, however as it will be further examined, the current methods used by ERT do not provide an accurate benchmarking environment. Apart from model construction, an important aspect of the CARM is the in-depth application characterization. 
Currently, Intel Advisor and AMD uProf provide application analysis, via DBI and PMUs respectively, while ERT only implements model generation (no support for application profiling/characterization). To close this gap in open-source cross-platform CARM-based application profiling, the developed CARM tool provides a robust application analysis subsystem alongside the CARM generation.

For in-depth architecture microbenchmarking and application analysis, it is typically required to access different performance counters of the diverse microarchitectures. For the architectures covered in this work, multiple state-of-the-art performance profiling tools were considered.  Consequently, the PAPI \cite{articlePAPI} tool was chosen to measure the PMU data during PMU analysis tests, due to its high level of portability and support for various architectures, which coincides with the objectives of this tool of keeping most of its features available on the most diverse set of architectures possible.

As an alternative to PMU profiling a DBI-based application analysis method was also implemented, DynamoRIO \cite{DynamoRIO} was selected as the main tool responsible for extracting an accurate opcode count used to then calculate the GFLOPS and memory traffic of a particular application or benchmark. DynamoRIO was chosen due to its compatibility with x86-64 and AARCH64 architectures, with RISC-V support currently undergoing its early stages, this set of supported architectures closely matches the architectures the CARM Tool intends to target, making it the ideal candidate for the DBI analysis support, additionally, Intel SDE \cite{SDE} DBI analysis was also included and supported.



%% file: CARM_Tool_high_level.tex
\begin{figure}[t!]
\centerline{\includegraphics[scale=0.6]{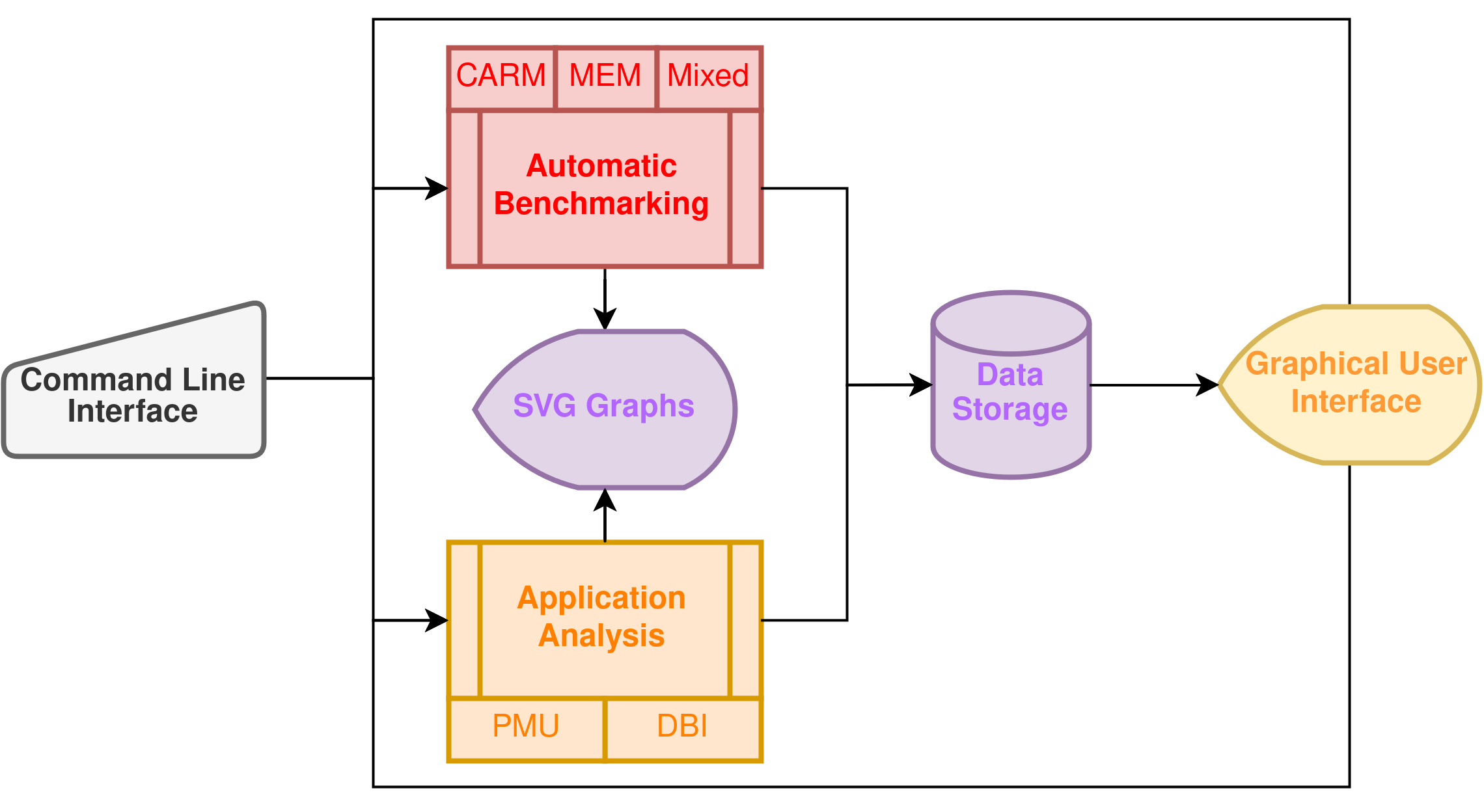}}
\caption{CARM tool modules}
\label{tool_high_level}
\end{figure}

The proposed CARM tool comprises a set of independent modules to provide a complete CARM-based profiling ecosystem as shown in Figure \ref{tool_high_level}. 
The tool offers both a command line and a graphical user interface (GUI) for interaction, allowing users to access stored results from benchmarking and application analysis. Whether activated through the command line or the GUI, the automatic benchmarking and application analysis modules execute user-specified tasks and automatically save the results, which can be visualized within the GUI or as SVG graphs.

\subsection{Automatic Benchmarking}

The \textit{Automatic Benchmarking} module in the CARM tool is managed through a Python script which coordinates automatic benchmark generation and execution. This module conducts various benchmarks to assess memory bandwidth and peak FP performance using automatically generated, tailored assembly level microbenchmarks across multiple supported microarchitectures. Benchmarks include roofline, multi-level memory bandwidth discovery (referred as memory curve benchmarks), and mixed-instruction types, which are detailed via microbenchmarks tailored to specific user-selected options and target specifications provided to the Python script via arguments (see below).

The '\textbf{--test}' argument specifies the benchmark type, such as '\textit{roofline}' for comprehensive CARM results or specific cache levels (\textit{'L1', 'L2', 'L3', 'DRAM'}) and \textit{FP} for focused tests. The '\textit{MEM}' option triggers memory curve benchmarks, while \textit{'mixedL1', 'mixedL2', 'mixedL3', and 'mixedDRAM'} target mixed benchmarks, which provide the ability to interleave the FP operations with memory accesses to a specific memory level. The '\textbf{--ISA}' argument selects ISA extensions for analysis, offering scalar and various vector options like SSE, AVX2 and AVX512 for x86-64, Neon for ARM, and RVV0.7/1.0 for RISC-V, while the default '\textit{auto}' option automatically detects and benchmarks available ISAs on the system. The data precision of benchmarks can be toggled between double and single precision FP data using the '\textbf{--precision}' argument, and the '\textbf{--threads}' argument allows specification of thread counts for multi-core benchmark execution.
The '\textbf{--ld\textunderscore st\textunderscore ratio}' argument serves to configure the ratio of load to store instructions present in a benchmark, while the '\textbf{--only\textunderscore ld}' and '\textbf{--only\textunderscore st}' flags can make the benchmark contain exclusively load or store instructions respectively, and allow for in-depth analysis of the bandwidth differences between load and store operations in various architectures.

\paragraph{Roofline Benchmarks}

The roofline benchmarks are central to the CARM tool, focusing on memory levels and two FP benchmarks. The first FP benchmark can be configured to use addition, division, or multiplication instructions via the '\textbf{--inst}' argument, while the second is always an FMA FP benchmark. For memory benchmarks, cache sizes are automatically detected on x86-64 systems using the \textit{cpuid} instruction, while they can be manually specified for AARCH64 or RISCV64 systems via a config file or command-line arguments.
The results of this benchmark are stored in a CSV file, and include the memory bandwidth of the various memory levels of the target system (in Gbps and IPC), and the peak FP performance for both FMA and add instructions by default (in GFLOPS and IPC). These results can be visualized in the GUI (see example in Figure~\ref{cara_mixed_final}). 

\paragraph{Memory Curve and Mixed Benchmarks}

Memory curve and mixed benchmarks are crucial components of the CARM tool, each providing unique insights into system performance upper-bounds. Memory curve benchmarks analyze variations in memory bandwidth across a wide range of problem sizes from 2Kb to 512Mb, and measure changes in bandwidth and memory IPC. This type of benchmark can be customized to use different load-store ratios, with results stored in a CSV file and visualized through an SVG graph. This graph, as shown in Figure \ref{quad_mem}, details memory bandwidth for various test sizes and can include cache size information to enhance data interpretation.
On the other hand, mixed benchmarks blend memory and FP benchmarks, targeting a specific memory level. Users can adjust the ratio of FP to load/store instructions  and select a specific FP instruction to use. The results, which include AI and GFLOPS metrics, are also recorded in a CSV file and can be viewed in the GUI (as dots on the CARM graphs, as illustrated in Figure~\ref{cara_mixed_final}). When  coupled, these benchmarks provide a complete view of system performance across different workloads and memory levels.

\subsection{Application Analysis}

As previousy referred, the proposed CARM tool provides an in-depth application profiling via two different subsytems: DBI and PMU.
DBI-based profiling is managed via two distinct facilities: a custom DynamoRIO client~\cite{DynamoRIO} or Intel SDE~\cite{SDE}. DynamoRIO can be used for reporting dynamic opcode counts on x86-64 and AARCH64, with support for RISCV64 in its early stages, while SDE is x86-64 only. Users must provide the executable path and select either DynamoRIO or Intel SDE for profiling, with installation paths specified accordingly.
DynamoRIO is available through direct download or compilation from source, and Intel SDE through its current release. 

The proposed CARM tool also incorporates specific profiling capabilities, which include \textit{Region of Interest (ROI)} code profiling. 
This functionality is facilitated by a header file which contains the API functions for the ROI instrumentation code (\textit{carm\textunderscore roi \textunderscore start(), carm\textunderscore roi \textunderscore end()}) necessary for ROI DBI application analysis.
The instrumented code measures execution times and categorizes opcodes by type and ISA for consultation, allowing for detailed FP and AI calculations, 
which are then stored in CSV files and can be visualized in the CARM Tool GUI (as seen in Figure \ref{fig:eigen_spmv}).

The PMU-based application profiling interfaces with the PAPI high-level API (\textit{PAPI\textunderscore hl\textunderscore region\textunderscore start(), PAPI\textunderscore hl\textunderscore region\textunderscore end()}), to calculate the AI and GFLOPS for profiled applications. This feature is currently exclusive for ROI profiling, with a possible future integration with tools like Likwid \cite{5599200} or Perf \cite{de2010new} to support full application profiling.
PAPI is configured to monitor events like `\texttt{PAPI\textunderscore LST\textunderscore INS}` (load/store instructions), `\texttt{PAPI\textunderscore SP\textunderscore OPS}` (single precision operations), and `\texttt{PAPI\textunderscore DP\textunderscore OPS}` (double precision operations), which are run three times to avoid PMU multiplexing issues. This setup ensures accurate profiling by avoiding the statistical assumptions required when sampling multiple events simultaneously. Profiling results can be consulted via the CARM Tool GUI just like with DBI profiling.

\subsection{Graphical User Interface}

\begin{figure}[t]
    \centering
    \includegraphics[width=1\columnwidth]{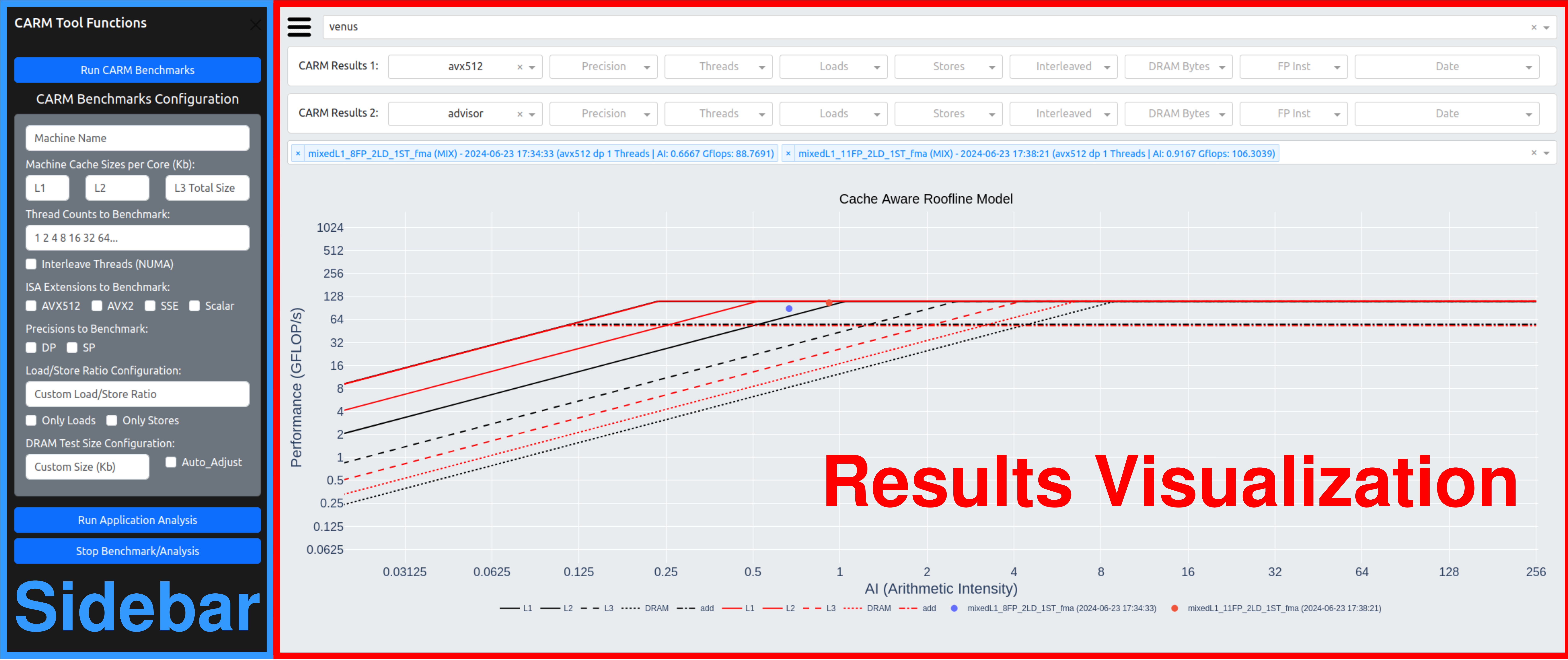}
    \caption{\small CARM Tool GUI Overview}
    \label{GUI}
\end{figure}

Although all functionalities can be accessed via command-line, the proposed CARM tool is also equipped with a browser-based GUI, shown in Figure \ref{GUI}, which facilitates the configuration  and execution of CARM benchmarks, application analysis, and the visualization of results. For this purpose, the GUI contains two main sections: \textit{i)} the main window responsible for the result visualization features (outlined in red); and \textit{ii)} the colapsable sidebar (outlined in blue) which implements various features of the CARM tool in an easy-to-use way, bypassing the need for the command line interface to obtain results. This GUI allows users to execute CARM benchmarks, profile applications, and visualize results in a convenient place, with an intuitive and easy-to-use interface.


%% file: CARM_Tool_low_level.tex
\begin{figure}[t!]
\centerline{\includegraphics[scale=0.7]{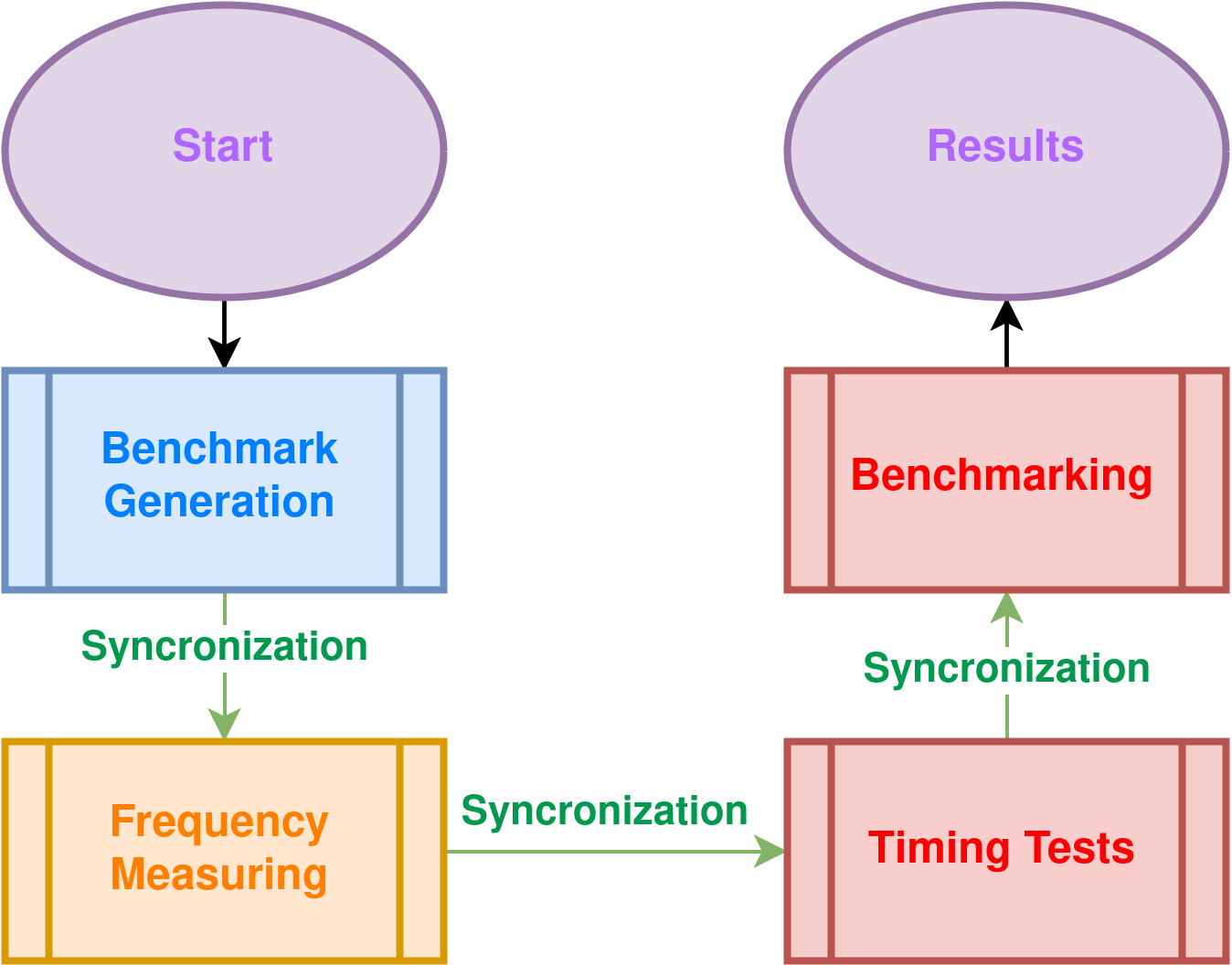}}
\caption{CARM tool microbenchmark execution steps}
\label{tool_low_level}
\end{figure}

The CARM tool microbenchmark generation and execution follows a methodology that contains a series of preparatory steps that are always performed on all supported architectures.
As outlined in Figure \ref{tool_low_level}, these steps include generating assembly level benchmarks based on user specifications and underlying ISA requirements, measuring the frequency of the target CPU, conducting timing tests to ensure the benchmarks run for an appropriate duration, and executing the benchmarking itself.
Besides the various steps, the necessary adaptations made on the automatic benchmark generation pipeline for each ISA are covered and explained in this section.

\subsection{Benchmark Generation}

\begin{figure}[!t]
\begin{lstlisting}[style=CStyle, caption={\small Memory microbench structure for X86-64 AVX2}, label=microbench_struct]]
__asm__ __volatile__ (
    "movq %0, %%r8"
    "Loop2_%=:"
    "movq %1, %%rax"
    "movq $16, %%rdi"
    "Loop1_%=:"
    "vmovapd 0(%%rax), %%ymm0"
    //...//
    "vmovapd %%ymm16, 8224(%%rax)"
    "addq $8256, %%rax"
    "subq $1, %%rdi"
    "jnz Loop1_%="
    //...//
    "subq $1, %%r8"
    "jnz Loop2_%="
    :"r"(num_reps_t),"r" (test_var)
)
\end{lstlisting}
\end{figure}

The automatic generation of assembly microbenchmarks for obtaining CARM results starts with configuring key parameters including the benchmark type, ISA, precision, load/store ratio, and data size for memory benchmarks (see Section~\ref{CARM_Tool_high_level}). These parameters instruct the \textit{Automatic Benchmarking} module to select the appropriate assembly opcodes and registers which are already automatically configured and defined for each of the supports ISAs when generating the desired microbenchmark.
Microbenchmarks are consistently built across all ISAs, following the structure shown in Listing~\ref{microbench_struct}. These include adaptable outer (\texttt{Loop2}) and inner (\texttt{Loop1}) loops, where the outer loop’s repetition rate is controlled externally to manage benchmark execution time (duration), while the inner loop’s number of iterations is predetermined to ensure it properly engages with the data for memory benchmarks or aligns with the expected number of FP operations for FP benchmarks. These settings are also used to address ISA-specific constraints affecting the number of iterations.




Typically, the inner loop executes 256 unrolled instructions that cycle through available registers to minimize data dependencies, encompassing a range of operations from FP arithmetic to load/stores of varying precision and ratios. Instructions exceeding this loop are added to the code remainder placed in the outer loop to maintain the required instruction count. In memory or mixed benchmarks, a designated variable holds an array pointer, which is accessed sequentially.
This benchmark structure is implemented in a controlled environment that prevents compiler optimization from omitting any essential instructions, ensuring that all benchmarks are executed as intended. Adaptations are made to this general structure to meet the specific needs and constraints of the supported ISAs.

\subsection{Frequency Measuring}
The frequency measuring step is crucial for obtaining IPC counts alongside timing during benchmark execution, allowing comparisons with the hardware's theoretical IPC limits. 
Simultaneous frequency measurement across all threads helps identify any fluctuations caused by different thread counts, ensuring reliable IPC counts for multithreaded executions.
During this process, assembly functions (one for each supported ISA) execute dependent scalar additions in a continuous loop (see the structure in Listing \ref{freq_measure_struct}), which inherently results in an IPC of one due to the data dependencies present in all operations. 
Since these functions run for a preset number of iterations, timed by the \textit{clockgettime()} function, the CPU frequency is calculated as a ratio between the known instruction count and the measured execution time.

\begin{figure}
\begin{lstlisting}[style=CStyle, caption={CARM tool frequency measuring assembly}, label=freq_measure_struct]]
clktest_loop:
    add %r8, %rbx
    //...//
    add %r8, %rbx
    sub %r9, %rdi
    jnz clktest_loop
ret
\end{lstlisting}
\end{figure}

\subsection{Timing Test and Benchmarking}

After frequency measuring, preliminary timing tests determine the optimal number of iterations to balance efficiency and result stability. The CARM tool automatically  adjusts the number of iterations in the outer loop of the microbenchmarking until an acceptable time duration is achieved, thus avoiding prohibitively short or long test durations and ensuring result reliability.
Actual microbenchmarking involves running the benchmarks 1024 times to maximize result accuracy and stability, coordinated with synchronization mechanisms like thread barriers to ensure uniform completion across threads, among other ISA-specific instructions such as \texttt{LFENCE} on x86-64 and \texttt{ISB} on AARCH64. For x86-64, the Time Stamp Counter (TSC) provides timing via cycle counts, while for AARCH64 and RISCV64, \textit{clockgettime} is used with nanosecond accuracy (to avoid the need for root privileges in some systems).
Each thread's best run per iteration is recorded, and the median of these runs determines the benchmark result, factoring in frequency and iteration count to accurately assess CARM metrics.
This methodology underpins the adaptations made for different architectures, ensuring each ISA's unique requirements and capabilities are appropriately addressed in the automatic benchmark generation and execution process.

\subsection{Benchmarking Architectural Adaptations}

While the general benchmarking structure, detailed in Listing \ref{microbench_struct}, is followed for all the supported architectures, special adaptations are made to properly extend microbenchmarking support across different ISAs and architectures. 

\subsubsection*{x86-64 Architectures}
On x86-64 architectures the previously elaborated microbenchmarking structure is strictly followed. However, for the timing of microbenchmarks (based on TSC), certain adjustments had to be made to reflect the fact that the cycle counts reported by the TSC are always based on the nominal CPU frequency (regardless of the currently set, real CPU frequency).
To address this, during the frequency measuring step on x86-64 systems the TSC is used alongside the \textit{clockgettime} function to count the cycles measured during the execution of the assembly function shown in Listing \ref{freq_measure_struct}. This approach allowed for the calculation of the CPU nominal frequency as a ratio between the obtained TSC cycles and the recorded time duration.
To properly scale the TSC cycle count to the actual frequency of the CPU and obtain the real elapsed cycles during the benchmark execution, a ratio of both frequencies is multiplied by the TSC cycle count, such that: 
\begin{equation} \label{realCycles}
RealCycles = TSC.Cycles \times \frac{RealFrequency}{NominalFrequency}.
\end{equation}

\subsubsection*{AARCH64 Architectures}
The AARCH64 ISA microbenchmarks had to be altered in order to take into account a limitation on the immediate values that can be used in AARCH64 assembly, which can not suprass the value of 4095. This is due to the RISC nature of AARCH64, which results in limited encoding space for immediate values in instructions. This limitation affects the number of repetitions in the inner loop of the microbenchmarks, thus no more than 4095 iterations can be performed in AARCH64. To resolve this issue, the CARM Tool automatically switches to a pointer-loaded variable for the inner loop control of AARCH64 benchmarks, whenever it is determined that more than 4095 iterations of the inner loop are necessary to achieve expected instruction counts, which bypasses the immediate issue.
Furthermore, the pointer offset of the memory array can not also be greather than 4095, meaning that the memory benchmarks must also take this into account, which is solved by limiting the amount of inner loop memory instructions to a maximum that does not allow the pointer offset to go beyond 4096. This reduction is then compensated by performing more inner loop repetitions or moving instruction to the outer loop to still perform the expected instruction counts.

\begin{figure}[!t]
\begin{lstlisting}[style=CStyle, caption={\small Assembly function to obtain max vector length}, label=riscv_vector_check]]
__asm__ __volatile__ (
    "li t0, 8192"
    "vsetvli t0, t0, e64, m1"
    "sw to, %[vl]"
    :[vl] "m" (vec_length)
    : "t0", "t1", "t2"
)
\end{lstlisting}
\end{figure}

\subsubsection*{RISC-V Architectures}
Similarly to AARCH64, the RISCV64 ISA also imposes limitation in the immediate offset of array pointers, which in this case is 2048, and it is overcome in a fashion similar to the one detailed for AARCH64. Other adaptations revolved around the integration of the RVV ISA extension in the microbenchmark structure. This is mainly because RVV poses a different challenge in regards to vector instructions, due to its vector size agnostic approach. 
This means that instructions can utilize a variable width/number of operands depending on what the underlying hardware can support, which brings to practice a new set of challenges specific for the RVV benchmarks. To perform vector instructions, first the desired vector length must be specified in the assembly code, using the \texttt{vsetvli} instruction. This instruction is used by the CARM tool before benchmark construction, as can be seen in Listing~\ref{riscv_vector_check}, to determine the maximum vector length available for the target CPU, and to automatically adjust the microbenchmarks to take advantage of the maximum vector length (or any valid vector length set by the user).


This function uses the \texttt{vsetvli} instruction to obtain the maximum vector length available. The \texttt{e64} specifies double-precision elements, so \texttt{t0} reflects the maximum number of such elements per vector, the \texttt{m1} field sets the register grouping to one vector register per instruction, in order to obtain only the elements of one vector. Larger groupings, up to \texttt{m8}, can handle eight vector registers simultaneously in each instruction for efficiency. 
This feature is particularly useful since the vector load and store instructions do not have space in their encoding for a pointer offset, meaning that the pointer must be bumped after every memory instruction. However with \texttt{m8} grouping, the negative effects on performance caused by this pointer bump can be minimized, by making it possible to only bump the pointer one time for every eight instructions.


%% file: experimental_results.tex
This section presents the experimental results of this work, covering the benchmark outcomes of the developed CARM and memory benchmarks within the CARM tool. It also details their validation through various means, including mixed benchmarks, DBI, and PMU analysis, all of which are capabilities of the CARM tool. Finally, it will provide a comparison with both Intel Advisor and ERT CARM analysis and an overview of the experimental results from a Sparse Matrix-Vector Multiplication (SpMV) application analysis. The various environments used for obtaining these experimental results are listed in Table \ref{machines}.

\subsection{CARM Benchmarking Results}

\subsubsection*{\textbf{x86-64 CPUs}}
\begin{figure*}[t]
    \centering
    \includegraphics[width=1\linewidth]{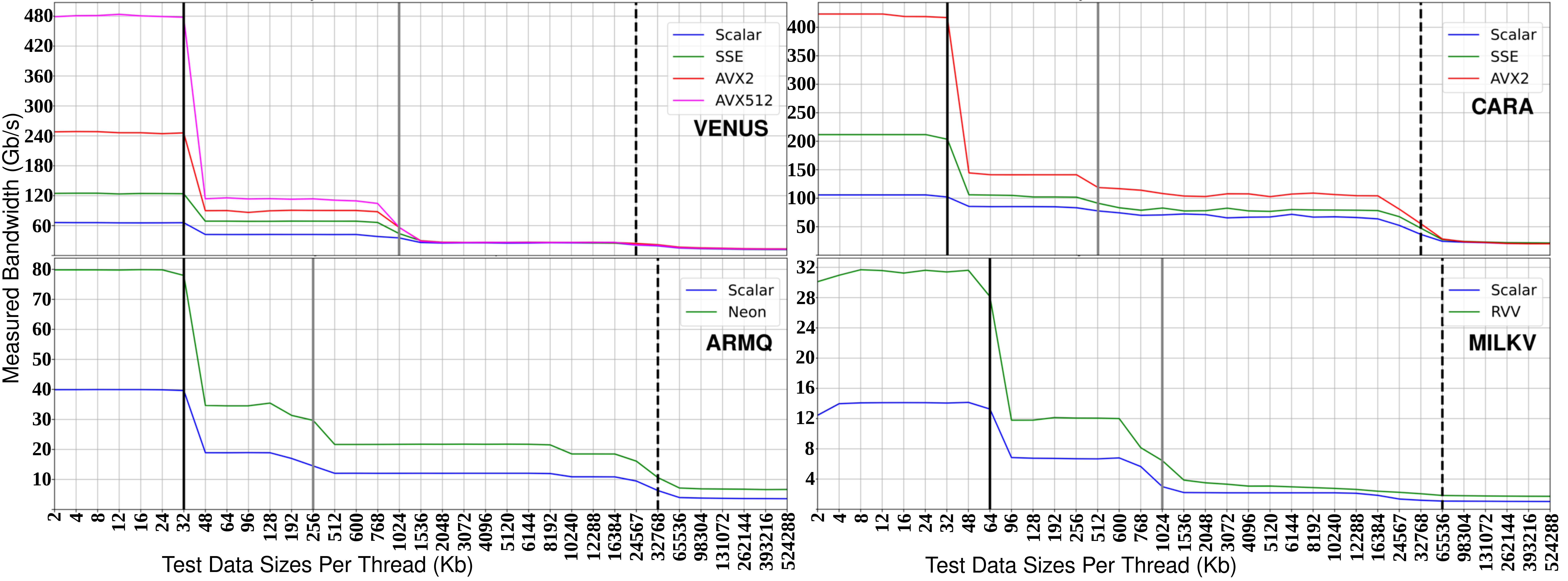}
    \caption{\small Memory curve benchmark results}
    \label{quad_mem}
\end{figure*}

For the x86-64 representation, the Venus and Cara machines were tested, as detailed in Table \ref{machines}, the Venus machine contains an Intel Skylake-X CPU which supports all x86-64 ISA extensions used by the CARM Tool, while the Cara machine contains an AMD Zen3 which supports up to AVX2. These CPUs, both feature two load and one store units per core, theoretically achieving up to three IPC at the L1 cache level. In order to better assess the complete performance of their memory subsystem, the memory curve benchmarks were conducted. These tests varied in load/store ratios, thread counts, and ISAs to measure their effects on bandwidth and IPC, as detailed in the memory curve graph for a 2 load to 1 store ratio on one thread for all available ISAs in Figure \ref{quad_mem}.

\begin{table}[t!]
\begin{center}
\caption{\label{machines}Machines used for experimental results}
\begin{tabular}{|c|c|c|} 
 \hline
 \textbf{Name} & \textbf{CPU} & \textbf{Architecture}\\
 \hline
 \textbf{Venus} & Xeon Gold 6140 & Skylake-X\\ 
 \hline
  \textbf{CSL} & Intel Xeon Gold 6258R & Cascade Lake\\ 
 \hline
 \textbf{Cara} & Threadripper PRO 5975WX & Zen3 \\
 \hline
 \textbf{Armq} & Cavium ThunderX2 CN9980 & Vulcan\\
 \hline
 \textbf{MilkV} & Sophon SG2042 & XuanTie C920\\
 \hline
\end{tabular}
\end{center}
\end{table}


From the various load/store ratios tests, it was observed that the two loads per store ratio produced the highest bandwidth and IPC values, matching the CPU's load/store unit ratio, hence this ratio is the one shown for the various ISA results. Load-only and store-only tests produced IPC counts of two and one respectively, which indicates the presence of two load and one store unit per core. On Venus however during the two loads per store test, the maximum theoretical IPC of three was not reached, only reaching a peak of 2.046 IPC. This value closely matches the cited value for "sustained bandwidth" by the Intel Optimization Manual \cite{IntelManual}, of 2.078 IPC, with only a 1.54\% deviation. For the L2 cache, while the manual cites a bandwidth of 0.813 IPC, our load-only benchmarks showed higher IPCs up to 0.964. L3 bandwidth analysis was complicated due to the small 1408Kb L3 slice per core. Tests with a 1216Kb data size, fitting between L2 and L3 limits, recorded a 0.194 IPC, 17\% below the reported 0.234 IPC for L3. 
The testing in the Cara machine yielded similar results, while there are no official sustained bandwidth values reported by AMD, the memory benchmarks managed to reach three, one, and 0.7 IPC for the L1, L2, and L3 memory levels, which closely match the theoretical architectural limits of the Zen3 architecture, with only the L3 result being 30\% below the theoretical maximum.

After analyzing the memory subsystem, we turn to the FP performance of the Skylake-X and Zen3 CPUs, which, with two FP units per core capable of AVX-512 and AVX2 operations respectively, can theoretically reach up to 2 FP IPC. To verify this the FP CARM benchmarks were executed for both AVX512 and Scalar on Venus, in this case, results showed IPC counts of 1.88 and 1.98 for AVX512 and scalar respectively, indicating that while not perfectly achievable we are able to get close to the theoretical 2 FP IPC, especially for the scalar instructions.
On Cara, two FP IPC were accurately reached, matching the expected theoretical maximums.
From this analysis, we can conclude that the CARM benchmarks on the x86-64 architectures are showing promising results, closely following the theoretical limits of the Skylake-X and Zen3 CPUs, these benchmarks will be validated in the next Section via various methods, to confirm that these measured values actually correspond to the real execution of the instructions in the assembly microbenchmarks ran.

\subsubsection*{\textbf{AARCH64 CPUs}}
The ThunderX2 CPU on the Armq machine, detailed in Table \ref{machines}, was also benchmarked. Equipped with the Neon SIMD extension, it has two Load/Store units capable of executing Neon operations, theoretically achieving up to two memory IPC. To explore these limits, memory curve benchmarks were conducted across various load/store ratios, in a similar fashion to the tests for x86-64. These benchmarks indicated that the load-only ratio performed the best on Armq, which is the ratio shown in the results that can be observed in Figure \ref{quad_mem}.


The load-only ratio benchmark reached an IPC of two, closely matching the expected IPC, using only stores resulted in an IPC of one, suggesting that only one of the load/store units in the CPU is capable of performing store operations. This discrepancy explains the stability of the load-only results, as introducing store instructions appears to cause conflicts within the dual-capable load/store unit. Furthermore, despite the ThunderX2 reportedly having a 64 bytes per cycle bandwidth to the L1 cache, suggesting a theoretical IPC of four, the practical IPC drops below one when accessing the L2 cache and decreases further with L3 cache usage.
The Armq CPU, equipped with two Neon-capable FP units per core, theoretically achieves an IPC of two for FP arithmetic. The FP CARM benchmarks, executed for both Neon and scalar extensions, show measured IPC values closely align with the theoretical two IPC, exhibiting an average deviation of 0.6883\% from this value.

\subsubsection*{\textbf{RISCV64 CPUs}}

The Sophon SG2042 CPU in the MilkV machine, contains a reported two load/store units per core, suggesting a theoretical IPC of two. Memory curve tests using the RVV extension indicate a peak IPC around one using a two loads per store ratio which was the best performing ratio, shown in Figure \ref{quad_mem}, suggesting the presence of only one effective load/store unit. Furthermore, store-only benchmarks consistently showed an IPC of about 0.5 across various data sizes, spanning from within the L1 cache limits to most of the L3 cache capacity. This trend is likely influenced by the C920 core's adaptive write-allocate policy \cite{c920Manual}, in conjunction with a 16Mb write buffer, which maintains steady bandwidth measurements across most data sizes. 
For FP performance, the MilkV machine, with two RVV-capable FP units per core, achieves an IPC close to two, as confirmed by CARM FP benchmarks with a minimal deviation (0.77\%) from the theoretical maximum for both RVV and Scalar instructions.

\subsection{CARM Benchmark Validation}
The validation of these benchmarks must be conducted to ensure the obtained results are indeed representative of the peak performance of the various machines, in this section, three different methods will be used to validate the benchmarks, all of which are available in the CARM Tool.

\subsubsection*{\textbf{Validation with Mixed Benchmarks}}
\begin{figure}[t]
    \centering
    \includegraphics[width=1\columnwidth]{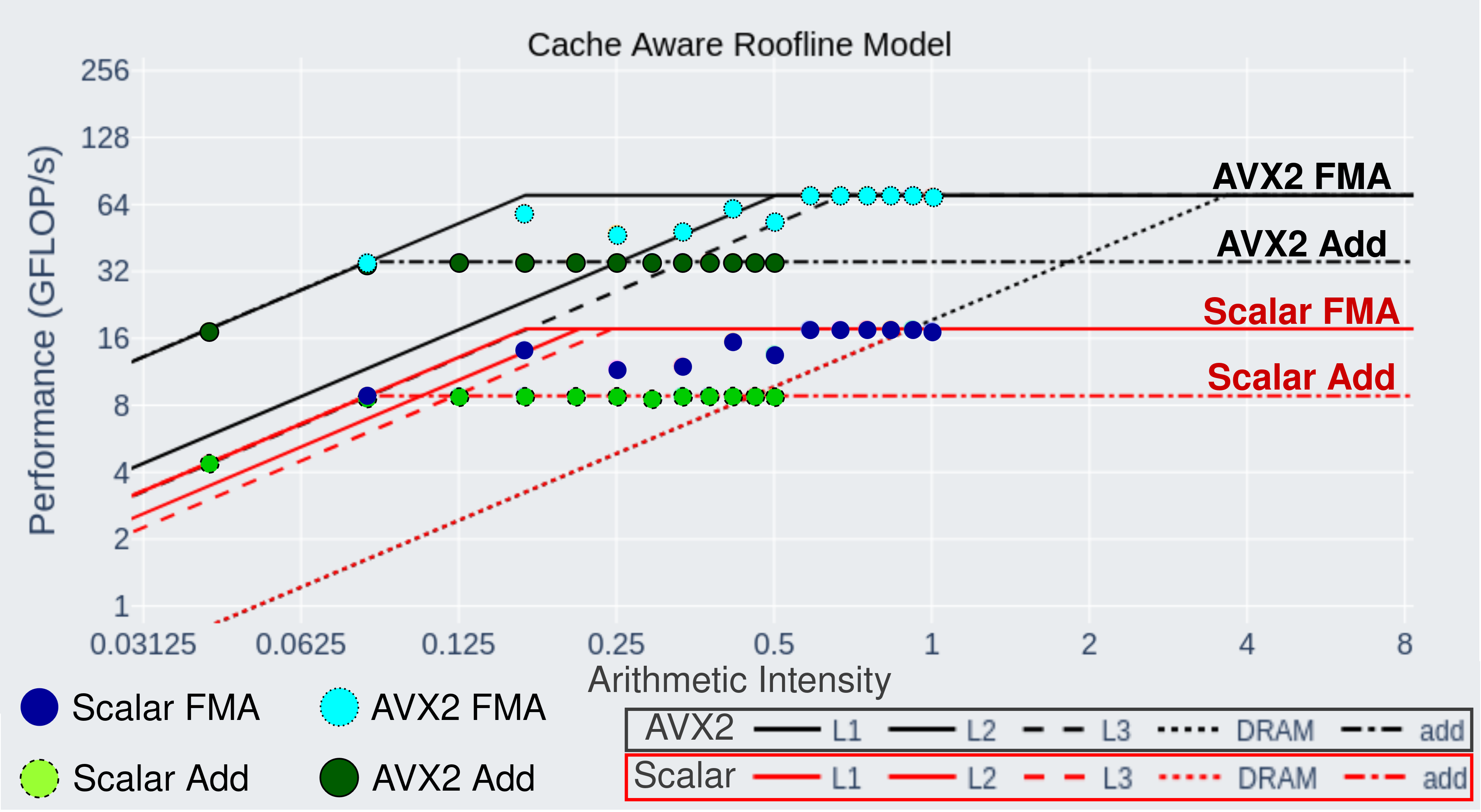}
    \caption{\small Cara mixed benchmark results}
    \label{cara_mixed_final}
\end{figure}

Examining the AMD Zen3 CPU on the Cara machine via the use of mixed benchmarks which allow users to adjust the FP to Load/Store instruction ratio of benchmarks that can target a specific memory level with a combination of memory and FP instructions. By using mixed benchmarks we can visualize the system performance when both the memory subsystem and FP units are stressed at the same time, which should still allow the CPU to reach near the limits set by the CARM benchmarking.
The Zen3 CPU with an optimal ratio of two loads per store, will execute various mixed benchmarks ranging from 0.0417 to 0.25 AI for addition (green dots) and 0.0833 to 0.5 AI for FMA instructions (blue dots). Mixed benchmarks targeting the L1 cache incrementally increased the FP instruction ratio to a maximum of 12 per three memory operations, capturing critical performance points around the ridge point of the CARM where FP limitations start to dominate.
These benchmarks, executed on the Cara machine using AVX2 and scalar instructions on one thread, showed that the Cara machine closely approached the CARM limits from previous CARM benchmarking, as shown in Figure \ref{cara_mixed_final} where each dot corresponds to a mixed benchmark execution of a particular AI. Errors in AVX2 were lower, averaging 13.69\% for FMA and 0.16\% for addition, compared to scalar benchmarks which showed errors of 13.97\% and 1.11\% respectively. While errors were considerably greater for FMA instructions in both cases, this suggests that less complex operations like addition maintain performance better under mixed conditions. 

\subsubsection*{\textbf{Validation with Dynamic Binary Instrumentation}}

While mixed benchmarks help validate CARM tool results, they don't confirm that the assembly instructions are executed as anticipated on the system. To ensure accuracy, we use the DBI analysis of the CARM Tool to analyze and verify opcode counts of the CARM benchmarks. By focusing the analysis on a ROI specifically targeting the benchmark execution sections, we minimize interference from other tasks, enhancing the precision of our measurements.
Additionally, to compare the accuracy and overhead of instrumented versus non-instrumented benchmarks, we maintain the same variables across tests and disable timing microbenchmark tests. We standardize the number of iterations in the benchmark’s outer loop, basing it on the initial runs of the non-instrumented benchmark to ensure consistency in test conditions, this method allows for direct comparisons between the two types of benchmarks.
Given the increased run-time due to DBI analysis, we reduce the number of benchmark repetitions from 1024 to 10. This adjustment is adequate because DBI, which simulates application execution while recording opcode counts, does not gain accuracy from higher repetition counts unlike methods that count directly during real execution, such as PMUs. This strategy ensures our benchmarks are both efficient and effective in validating assembly instruction execution.

Continuing with this established methodology, the L1, L2, L3, and FP benchmarks were executed on the Armq machine, focusing exclusively on the load-only ratio, which was found to perform best on this system from previous CARM benchmarking. Table \ref{dynamoBenchValidationARMQ} presents a comparison of expected versus measured instruction counts using DynamoRIO for these benchmarks. 
The DBI analysis on the Armq system showed only slight discrepancies, with deviations ranging from 0.36\% to 2.08\%. Despite these variations, the observed time overhead was minimal, barely noticeable amid the expected timing fluctuations of the benchmark, particularly given the reduced repetition count of 10, in contrast to the typical 1024 repetitions in standard benchmarks. 

\begin{table}[t!]
\begin{center}
\caption{\small Expected instruction counts vs measured instruction counts}
\label{dynamoBenchValidationARMQ}
\begin{tabular}{|c|c|c|c|c|} 
 \hline
 \textbf{Test} & \textbf{Loads} & \textbf{DR Loads} & \textbf{FP} & \textbf{DR FP}\\
 \hline
 \textbf{L1} & 4830192640 & 4887081032 & X & X\\ 
 \hline
 \textbf{L2} & 2144542720 & 2190049352 & X & X\\
 \hline
 \textbf{L3} & 1109606400 & 1115849352 & X & X\\
 \hline
 \textbf{FP ADD} & X & X & 4995153920 & 5013340160\\
 \hline
\end{tabular}
\end{center}
\end{table}

\subsubsection*{\textbf{Validation with Performance Counters}}

\begin{figure}[!t]
    \centering
    \includegraphics[width=1\columnwidth]{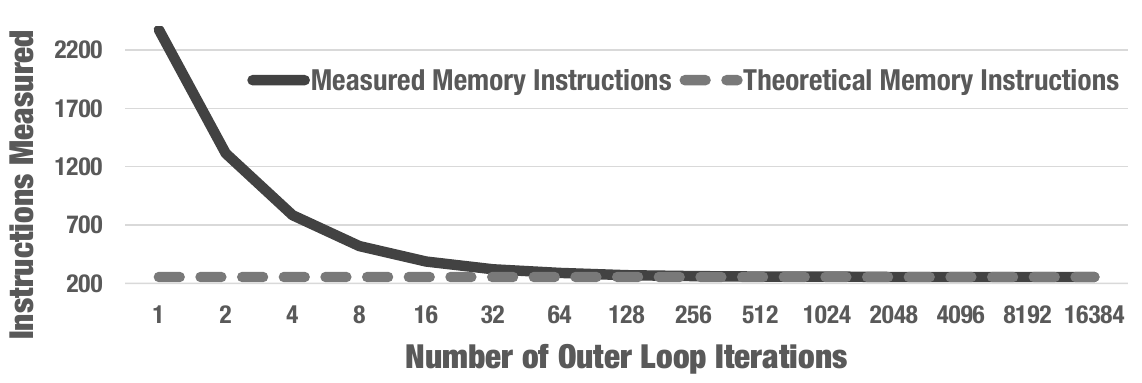}
    \caption{\small PMU analysis results}
    \label{fig:dynamo_mixed_venus_mem}
\end{figure}

Performance counters, similarly to DBI, can be used to confirm the execution of instructions in benchmarks, in this case, profiling focused on assessing the overhead of performance counter instrumentation in counting FP and memory instructions, which was performed on the Venus system. A mixed benchmark employing the AVX512 ISA at targeting the L1 memory level was adjusted to perform only one iteration of the outer loop and executed a single time to highlight any potential measurement overhead. The outer loop's iterations were then progressively increased to track deviations between the measured and actual expected instruction counts. The expected benchmark values, 255 memory instructions, and 2040 FP operations, were then compared against the PMU-measured values as the loop iterations increased, as illustrated in Figure \ref{fig:dynamo_mixed_venus_mem} for the memory instructions. Notably, deviations in memory instruction counts were significantly less accurate than those in FP operations for small iteration counts, achieving 1.6\% deviation only after 512 iterations, while FP counts showed only a 0.3\% deviation with only one iteration of the outer loop.

\subsection{Intel Advisor Analysis and Comparison}
\begin{figure}[t]
    \centering
    \includegraphics[width=1\columnwidth]{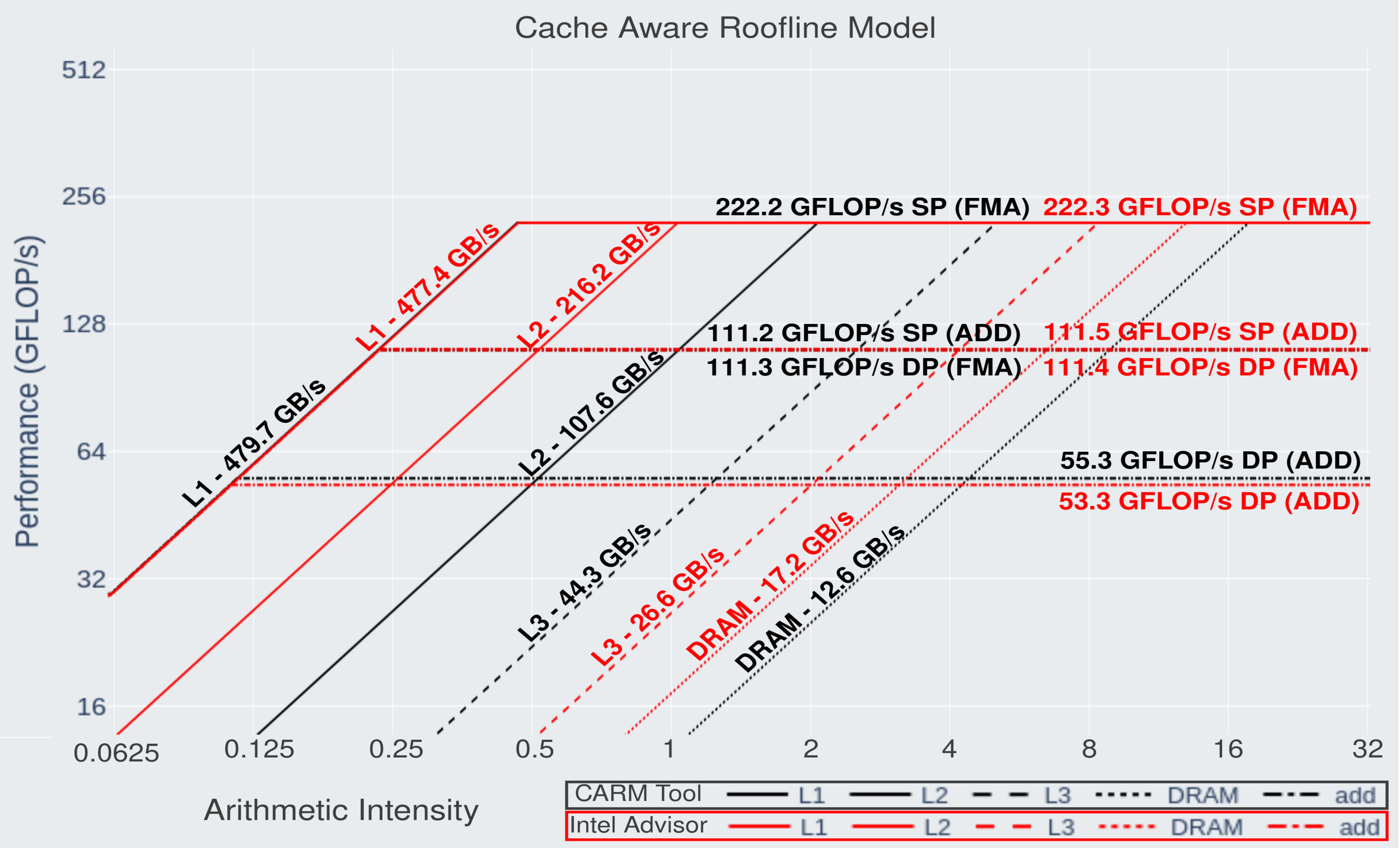}
    \caption{\small Intel Advisor and CARM Tool Venus AVX512 results}
    \label{fig:adv_vs_carm}
\end{figure}

Intel Advisor implements the CARM for Intel CPUs, measuring the peak performance of the various memory levels and FP arithmetic instructions using the AVX512 ISA for single and double precision. In this comparison, the Venus machine was used to run Intel Advisor and obtain the AVX512 roofline data from Intel Advisor (in red) as can be seen in Figure \ref{fig:adv_vs_carm}, compared with the CARM Tool results (in black) using a two-load-per-store ratio on the memory benchmarks. Since Intel Advisor also tests the single precision arithmetic FP performance of AVX512 by default, these lines were also included in the CARM graph for comparison.

As can be observed in Figure \ref{fig:adv_vs_carm}, the obtained L1 bandwidth and FP roofs closely match on both tools with a variation of only 0.48\% between measured values. 
In the case of the L2 bandwidth, Intel Advisor presents a considerably higher measurement, which can be explained by the usage of a different load/store ratio by Intel Advisor for the L2 test used in order to achieve the highest possible bandwidth. By performing the L2 bandwidth test using the CARM Tool with a load-only ratio instead of the two-loads-per-store ratio, the CARM Tool is able to achieve up to 221.11 GB/s, which surpasses Intel Advisor's measurement by about 2.46\%. 
On the other hand, the CARM Tool achieves a much higher bandwidth value for the L3 cache, which is due to the CARM Tool using a data size of 1216Kb meant to fit inside the L3 slice of a single core in a Skylake-X system, while Intel Advisor could be considering a data size that is meant to fit in the entirety of L3 while not fitting inside an individual core's slice (not possible to assess due to the closed-source nature of Intel Advisor). However, if we consider a different data size for the L3 benchmark of the CARM tool, such as half of the L3 cache size (12.88Mb), in which case the CARM Tool obtains a bandwidth value of 26.45 GB/s, much closer to Intel Advisor's measurement. 
In regards to the DRAM bandwidth, Intel Advisor obtains a higher bandwidth in this test, which can still be atributed to the varying load/store ratio used in Intel Advisor, e.g., the CARM Tool result (12.6 GB/s) comes from a 512Mb array with two loads per store, while a higher value of 19.95 GB/s can also be obtained by using a different array size (50Mb) and only loads, surpassing the measurements of Intel Advisor.

\subsection{ERT Analysis and Comparison}
\begin{figure}[t]
    \centering
    \includegraphics[width=1\columnwidth]{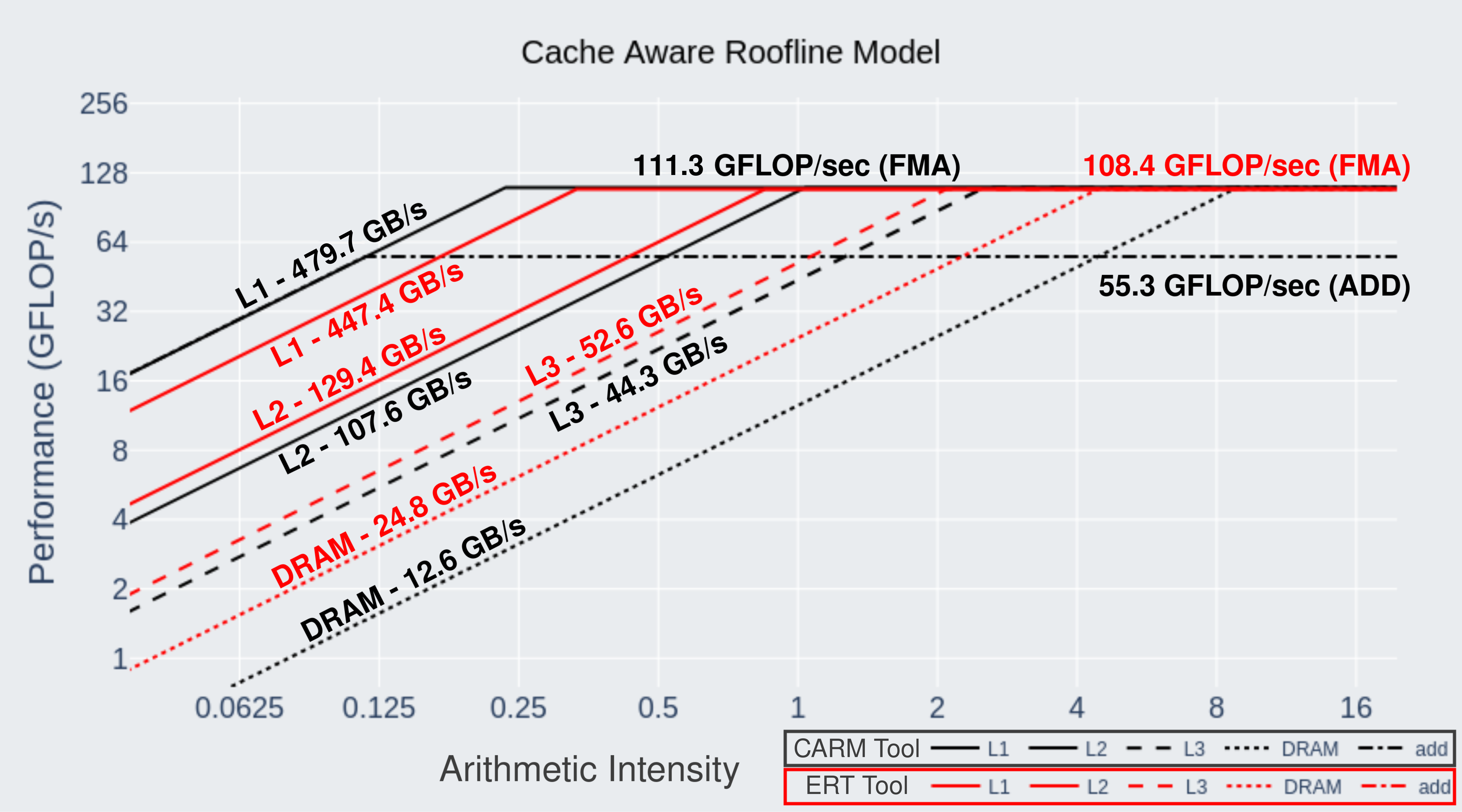}
    \caption{\small ERT and CARM Tool Venus AVX512 results}
    \label{fig:ert_vs_carm}
\end{figure}

As mentioned the ERT provides an open-source implementation of the roofline model, although which version of roofline it adapts is not explicitly mentioned, it produces the necessary data for a comparison with the CARM generated by the CARM Tool. 
For this comparison, the Venus machine was used to obtain the peak FP and memory level bandwidths from ERT, using the AVX512 instructions set. These values will now be subsequently compared with the ones from the CARM tool, and observations on the implementation details of ERT will be made. 
Figure \ref{fig:ert_vs_carm} showcases two CARM graphs, one from the CARM tool (in black) and another from ERT (in red). As can be observed, the CARM tool is able to obtain a higher L1 bandwidth and a slightly higher GFLOPS roof. However, the ERT results show higher bandwidth values for all of the other lower memory levels. When comparing these bandwidths with the load-only results obtained for the Intel Advisor comparison, the CARM Tool is able to surpass the L2 bandwidth values obtained by ERT and approach the DRAM measurements.
The main reason for this discrepancy lies in how the ERT tool determines cache sizes, and which test size results are considered for each memory level. Essentially the ERT tool performs a series of memory benchmarks with increasing problem sizes while measuring bandwidth levels via timing of a kernel with pre-determined amounts of memory instructions. These bandwidth results are then statistically smoothed to reduce irregularities, then the ERT tool determines the existence of a new memory level based on the variation between bandwidths across similar test sizes. If there is enough of a difference in bandwidth between close test sizes, ERT will assume a new memory level has been reached and then consider the following bandwidth values for the total bandwidth calculation of that new memory level. This resulted in tests sometimes showing more than three cache levels, indicating this method can be prone to inaccuracies.
Furthermore, since ERT does not provide a clear way to verify what test sizes were used for the calculation of which memory bandwidth level, and no way of manually specifying these values, it becomes impossible to determine how accurate it is in comparison with the real cache sizes. This likely explains the higher bandwidth levels of all the lower cache levels, since they most likely include some influence of the levels above them.



\subsection{SpMV Cross-Architecture Analysis}

An application analysis study was run on an SpMV implementation using the Eigen library \cite{EigenLibrary} in order to showcase the application profiling features of the CARM Tool on a real application use case. For this, the performance benefits of using a Reverse Cuthill McKee (RCM) \cite{10.1145/800195.805928} re-ordered matrix for SpMV execution were evaluated using the Eigen library on both x86-64 and AARCH64, while also providing a look at the differences in performance of the Eigen SpMV implementation in these distinct architectures.

\begin{figure}[t]
    \centering
    \includegraphics[width=1\columnwidth]{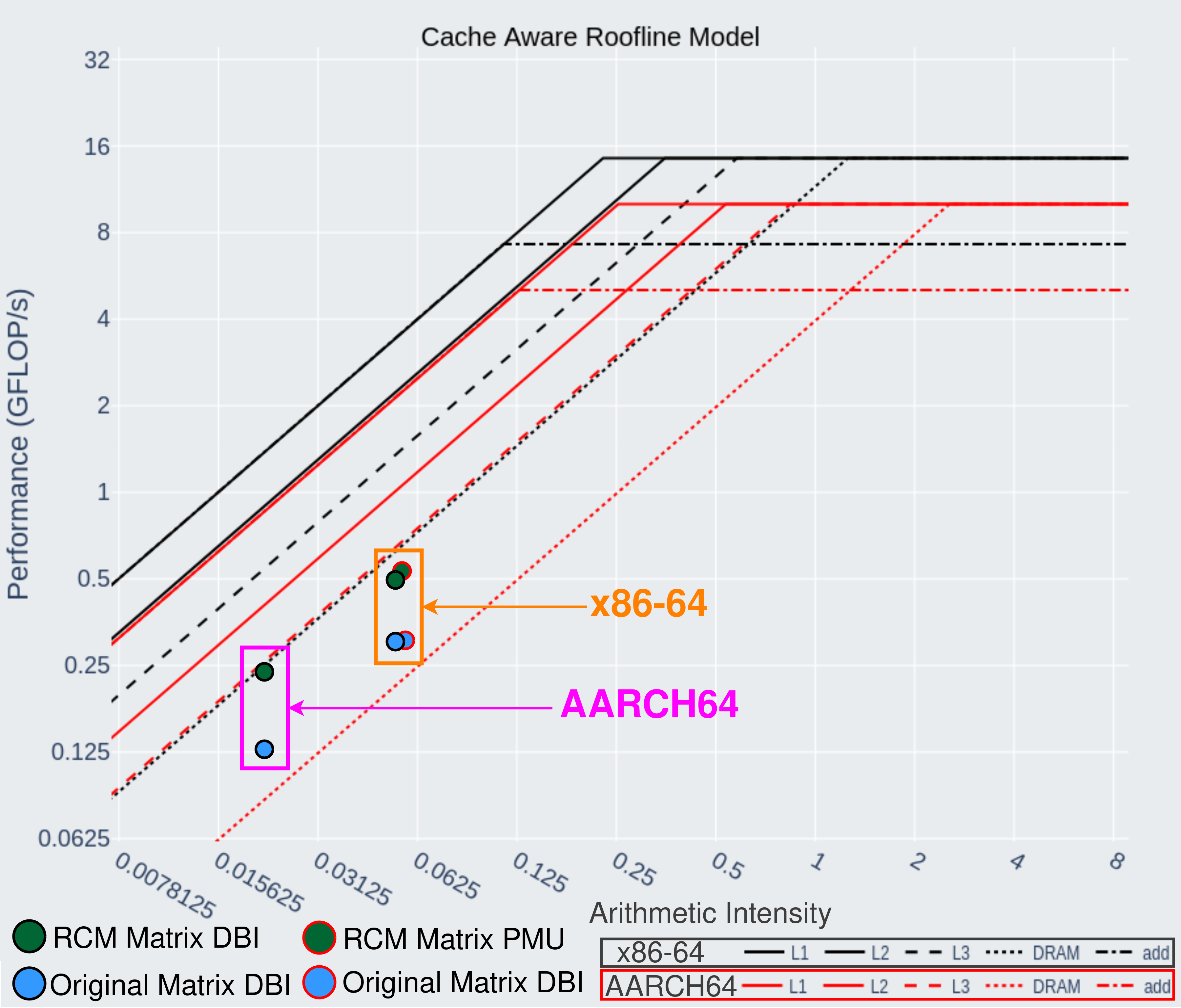}
    \caption{\small CARM results of SpMV analysis}
    \label{fig:eigen_spmv}
\end{figure}

The results of this application analysis using the \texttt{hugetrace-00020}\cite{sparse-tamu} matrix can be seen in Figure \ref{fig:eigen_spmv}, in this CARM plot we can observe the scalar single-threaded CARM results of both Venus (x86-64) and Armq (AARCH64), it can be noted that the Venus machine has better overall FP and memory bandwidth performance, which contributes to the higher observed GFLOPS values. The dots in the plot represent the SpMV runs executed, with the blue and green center indicating a run using the RCM and original matrix respectively, 
due to restrictions in PMU access on Armq, this analysis was only carried out on Venus, and it is represented by the red outline on the dots, while the black outline indicates DBI analysis. For Armq DynamoRIO was used to perform DBI instrumentation, while on Venus Intel SDE was used. On both machines the effects of RCM re-ordering are evident, while keeping the same AI since the underlying algorithm does not change, the reported GFLOPS increased by about 53\% on Armq and 59\% on Venus. 
Furthermore, we can observe that the reported values by the PMU analysis and DBI analysis on Venus almost coincided, with the PMU analysis reporting only 4.04\% more GFLOPS and a 5.26\% higher AI than the DBI on average.
Finally, the positioning of all the runs in the CARM graph indicates a clear memory bottleneck, which goes in line with what is expected from an SpMV implementation. This study showcases how the CARM graph can be leveraged alongside the application profiling capabilities of the CARM tool to conduct intuitive application performance analysis and bottleneck identification.




%% file: conclusion.tex
Continuous micro-architectural enhancements have significantly increased the complexity of modern systems' underlying hardware. 
As we approach the era of exascale computing, systems are expected to incorporate thousands of cores and integrate advanced technologies across various hardware subsystems. 
This complexity poses significant challenges for application designers striving to maximize the performance potential of these sophisticated machines. Consequently, robust and insightful methods like the CARM are indispensable for rapidly analyzing application bottlenecks and providing optimization guidance.
This paper addressed the challenge of creating an automated performance modeling tool tailored to the complexities of diverse HPC systems,
via the development of the CARM Tool, a versatile platform that supports CPUs from major manufacturers like Intel, AMD, ARM, and RISC-V.
Additionally, the inclusion of a GUI simplifies interaction with the CARM model, making it accessible even to users unfamiliar with complex performance modeling. This GUI allows users to effortlessly navigate and utilize the tool according to their specific requirements.
In essence, this article not only introduced a valuable tool but also enriched our comprehension of measuring and enhancing HPC system performance. 
This tool stands to significantly aid in keeping pace with the evolving demands of the computing world, empowering users to optimize their systems effectively.
This work establishes a robust foundation for future research and practical implementations in HPC system optimization and facilitates advanced performance optimization for a broad audience.

%% file: artifact_appendix.tex
\subsection{Abstract}

In recent years, HPC systems and CPU architectures as their central components have become increasingly complex, making application development and optimization quite challenging. To this respect, intuitive performance models like the Cache-aware Roofline Model (CARM) offer effective guidance by providing insights into bottlenecks that limit the application's ability to reach the system's maximum performance. To fully exploit the benefits of CARM optimization guidance for application development, automatic tools for cross-architecture model construction and in-depth application characterization are absolutely essential. Given a plethora of existing CPU architectures, the current landscape of CARM-enabled tools covers either vendor-specific (Intel Advisor), not sufficiently developed (ARM) or simply non-existing (AMD, RISC-V) tools. This is a particular gap that this work intends to close by bringing the automatic CARM support to all major CPU architectures and ISAs, i.e., x86 (Intel, AMD), ARM, and RISC-V, by developing assembly microbenchmarks specifically tailored to cover a full performance spectrum of modern CPUs (from scalar to all supported vector ISA extensions) for both computational units and all memory hierarchy levels. Additionally, this work integrates application analysis within the CARM framework using performance counters and dynamic binary instrumentation. Experimental results from this automated CARM framework show less than a 1\% deviation across various tested architectural maximums.

\subsection{Artifact check-list (meta-information)}


{\small
\begin{itemize}
  \item {\bf Program: } PAPI, Intel SDE, DynamoRIO
  \item {\bf Compilation: }
  GCC by default
  \item {\bf Run-time environment: } Linux
  \item {\bf Hardware: }
  x86-64, AARCH64, RISCV64
  \item {\bf Execution: } Command line or GUI
  \item {\bf Metrics: }
  Arithmetic Intensity, GFLOPS
  \item {\bf Output: }
  SVG graphs, CSV files, browser interface
  \item {\bf Experiments: Benchmark execution and Application Analysis}
  \item {\bf How much disk space required?: }
  Under 1Gb
  \item {\bf How much time is needed to prepare workflow?: }
  Under 1 hour
  \item {\bf How much time is needed to complete experiments?: }
  Around 4 hours or less
  \item {\bf Publicly available?: }
  Yes
  \item {\bf Code licenses (if publicly available)?: }
  LGPL
  \item {\bf Archived (provide DOI)?: } 10.5281/zenodo.12805280
\end{itemize}
}

\subsection{Description}
The artifact consists of a tool capable of performing the necessary benchmarks to generate the Cache-Aware Roofline Model (CARM) for CPUs of various architectures (x86-64, AARCH64, RISCV64), via a command line or graphical user interface (GUI). Furthermore, the tool allows for application analysis in the context of the CARM model using Performance Monitoring Units (PMUs) or Dynamic Binary Instrumentation (DBI).

\subsubsection{How to access}

The tool can be accessed via its \href{https://github.com/champ-hub/carm-roofline}{GitHub repository}. Users need to clone this repository for the artifact evaluation.

\subsubsection{Hardware dependencies}
A system that contains an x86-64 CPU (Intel Skylake-X) is ideal for reproducing most results from the paper, however, other CPUs can be used, in which case an AVX-512 capable CPU allows for more comparable results.

\subsubsection{Software dependencies}
The tool has been mostly tested under Linux Ubuntu or Cent OS, however, any Linux distribution should also work.

\noindent
For the tool itself, to generate SVG memory curve graphs the following Python packages are required:
              plotly;
              numpy;

\noindent
For the Graphical User Interface some form of browser is required and the following Python packages:
              dash; 
              dash-bootstrap-components;
              plotly;
              numpy;
              pandas; 
              diskcache;

\noindent
For the application analysis functionalities, PAPI, and Intel SDE (or DynamoRIO) are required to enable PMU and DBI-based analysis respectively.


\subsection{Installation}


After cloning the repository and installing the necessary Python packages, no further steps are required to obtain the CARM, memory curve, and mixed benchmark results for the target system. 
To enable the application profiling facilities, the PAPI and Intel SDE (for x86-64 only or DynamoRIO for x86-64 / AARCH64 DBI) tools must also be installed. We recommend installing PAPI following this \href{https://github.com/icl-utk-edu/papi/wiki/Downloading-and-Installing-PAPI}{guide}. The Intel SDE tool can be downloaded from its \href{https://www.intel.com/content/www/us/en/download/684897/intel-software-development-emulator.html}{website}.
A release of the DynamoRIO tool can be \href{https://github.com/DynamoRIO/dynamorio/releases}{downloaded} with no extra steps required.

\subsection{Experiment workflow}

To reproduce the AVX-512 CARM 
results presented 
for the Skylake-X machine,
users must run the command:
\begin{lstlisting}[style=DOS]
python3 run.py --isa avx512 -v 3
\end{lstlisting}

\noindent
The tool should be able to automatically detect cache sizes and confirm the availability of the AVX-512 ISA vector extension in the system if present, the \textbf{-v 3} argument is used to obtain a more detailed intermediary output of the benchmarks which can contain relevant metrics such as the peak bandwidth of each memory level. This process can also be done using the GUI, by running the \textbf{ResultGUI.py} script, 
then users can execute the benchmarks in their local system by using the sidebar to configure the benchmarking details.
, in this case, the AVX-512 box must be checked or all vector ISA extensions will be tested by default when clicking the "\textbf{Run CARM Benchmarks}" button. 
The CARM results can be found stored in a CSV file inside the ./Results/Roofline folder. In case tests are run on a machine without a browser available, we recommend moving the obtained results to a machine that has a browser to greatly improve the results visualization experience via the GUI.

In order to obtain similar memory curve graphs to the ones presented in the paper users must run the command:
\begin{lstlisting}[style=DOS]
python3 run.py --isa avx512 --test MEM --plot -v 3
\end{lstlisting}
\noindent
This command will perform the necessary benchmarks to obtain the memory curve graph for the AVX-512 ISA on the underlying system. These results can be consulted in the SVG file generated in the "\textit{./Results/MemoryCurve folder}". To vary the load/store ratio from the 2-1 default, the "\textbf{-ldst $<$load-to-store-ratio$>$}" argument can be used, or to perform exclusively loads or stores the "\textbf{-only\textunderscore ld}" and "\textbf{-only\textunderscore st}" arguments can be used.

To obtain the mixed benchmark results that can be then compared with the CARM results previously obtained as was done in the paper the following command must be used:
\begin{lstlisting}[style=DOS]
python3 run.py --isa avx512 --test mixedL1 -fpldst 1 -v 3
\end{lstlisting}

\noindent
In this command the "\textbf{--inst fma}" flag can be inserted to perform the mixed benchmarking using FMA instructions (the default is add). Furthermore, to vary the ratio of FP instruction in relation to memory instructions the "\textbf{--fpldst}" argument can be used.

To perform the application analysis of the same SpMV implementation used in the paper, the Intel SDE (or DynamoRIO) and PAPI tools must also be present in the system. The SpMV implementation showcased in the paper is also made available in the repository inside the SpMV folder. However, this application requires the presence of the Eigen library, which can be downloaded from its \href{ttps://gitlab.com/libeigen/eigen/-/releases/3.4.0}{repository}, and then extracted and inserted in the SpMV directory, with no further steps required.
Then, compile the SpMV application using the commands:
\begin{lstlisting}[style=DOS]
g++ -std=c++11 -O3 -mavx -mfma -I/${PAPI_DIR}/include -L/${PAPI_DIR}/lib -I./eigen-3.4.0 -o eigen_spmv_PMU eigen_spmv_PMU.cpp -lpapi

g++ -std=c++11 -O3 -mavx -mfma -I./eigen-3.4.0 -o eigen_spmv_DBI eigen_spmv_DBI.cpp
\end{lstlisting}

\noindent
To obtain the executable that can be analyzed using the PMU and DBI methods of the CARM Tool. 
After generating both executables the applications can then be analyzed using the CARM Tool with the same matrices used in the paper, (which must first be extracted from their archive and placed in the SpMV directory) using the commands:
\begin{lstlisting}[style=DOS]
python3 PMU_AI_Calculator.py /path/to/PAPI ./SpMV/eigen_spmv_PMU ./SpMV/hugetrace-00020_rcm.mtx <number of iteration>

python3 DBI_AI_Calculator.py --roi --sde /path/to/SDE ./SpMV/eigen_spmv_DBI ./SpMV/hugetrace-00020.mtx <number of iteration>
\end{lstlisting}

\noindent
These commands will respectively run the CARM Tool analysis of the SpMV application using both performance counters and DBI, these results can then be consulted in a CARM graph using the GUI. The analysis can also be executed via the GUI, by clicking the "\textbf{Run Application Analysis}" button in the sidebar, however in this case DynamoRIO must be used since support for SDE in the GUI is still a work in progress.

\subsection{Evaluation and expected results}


From the CARM roof results reviewers can expect similar GFLOPS and bandwidth lines as the ones presented for the Venus machine in the paper if also using a Skylake-X CPU. Most CPUs will tend to have between 3 and 2 load/store units per core meaning that instruction per cycle (IPC) values of 2 are expected in most cases when it comes to the L1 memory tests, with a downward trend as higher memory levels are tested. On the FP arithmetic side, most CPUs have 2 FP units per core, meaning that 2 IPC are expected in the FP benchmarks. To confirm these theoretical limit values reviewers should use the \textbf{-v 3} flag to more conveniently observe these measurements.

For the memory curve graph, there should be a clear drop in reported bandwidth as the problem size used by the benchmarks is increased, especially after surpassing the size of each memory level. There should also be some significant variations if doing tests with only stores, and sometimes even with only loads (this also applies to the CARM, and mixed benchmark results).

From the mixed benchmark validation, reviewers can expect the reported dots to come close to the upper CARM roofs of the same system, with mixed benchmarks using the addition instruction reaching near the first flat CARM roof, and the higher flat CARM roof when FMA instructions are used.

Finally from the application analysis, reviewers can expect similar reports in AI and GFLOPS from both the performance counter and DBI-based analysis methods, and also an increase in reported GFLOPS when running the analysis using the provided RCM re-ordered matrix when compared with the original matrix. Furthermore, the SpMV application will also list a theoretical number of flops performed (based on the number of non-zeros of the matrix) and its own execution timing results, which can be compared with the ones obtained from the analysis results.

\subsection{Experiment customization}

Reviewers can take full advantage of the different benchmark customization options of the provided tool that can be used to vary the parameters used for the benchmarks executed,
which are fully documented in the README of the tool, and on its \href{https://champ-hub.github.io/projects/The_CARM_Tool/}{website} page.

\subsection{Notes}

We also encourage reviewers to compare the CARM Tool results with other prominent proprietary tools that implement the CARM model such as Intel Advisor, in this case, we must note that Intel Advisor will vary the load/store ratio during its testing of the different memory levels to produce the best result possible, while the CARM Tool follows the load/store ratio defined by the user for all the memory levels, which can result in some lower (or higher) bandwidth levels when compared with Intel Advisor. For this reason, we recommend varying the load/store ratio, especially by running load-only tests which, as we have observed for some systems, can obtain better performance in memory levels higher than L1 (on Intel Skylake-X for example) and should be able to match the performance obtained by Intel Advisor on the same system.